\documentclass[twocolumn,showpacs,preprintnumbers,amsmath,amssymb]{revtex4}

\newcommand{\integ}[1]{\int d#1 \hspace*{1mm}}

\usepackage{graphicx}
\usepackage{dcolumn}
\usepackage{bm}
\usepackage[dvips]{color} 

\begin{document}

\preprint{}

\title{Polaron formation as a genuine nonequilibrium phenomenon}

\author{Shigeru Ajisaka} \email{g00k0056@suou.waseda.jp}
\author{Shuichi Tasaki}%
\author{Ichiro Terasaki}%
\affiliation{%
Department of Applied Physics, School of Science and Engineering,
Waseda University, Tokyo 169-8555, Japan
}%

\date{\today}

\begin{abstract}
Solitons and polarons at nonequilibrium steady states are investigated for the spinless Takayama Lin-Liu Maki model. Polarons are found to be possible {\it only out of equilibrium}.
This polaron formation is a
genuine nonequilibrium phenomenon, as there is a lower threshold current below which they cannot exist.
It is considered to be an example of microscopic dissipative structure.

\end{abstract}


\maketitle
Nearly 30 years ago, the nonmetal-metal transition by doping halogens in polyacetylene
was discovered by Shirakawa {\it et.al.}~\cite{Shirakawa}. As a result of observing a mobile defect in polyacetylene, Su, Schrieffer and Heeger~\cite{SSH1} studied solitons within
a lattice model (SSH model). Then, Takayama, Lin-Liu, and Maki~\cite{TLM} showed that
its continuum version, the so-called TLM model, allows one to study solitons analytically.
Subsequently, polarons were numerically found in the SSH model~\cite{SS}, and 
Brazovskii-Kirova\cite{BzCB1} and Campbell-Bishop \cite{BzCB2} have independently 
demonstrated the existence of polarons within the TLM model.
Subsequently, its electronic and transport properties have been extensively studied both 
experimentally and theoretically~\cite{SSHrev}, and it is known that the charge carriers 
in conjugated polymers are localized excitations such as solitons and polarons. 
In spite of these developments, there still remain several important issues to be settled.
Recently, the critical dopant concentration for the nonmetal-metal transition was improved~\cite{Heron}
and the absence of the soliton contribution to the current was reported~\cite{Leonardo}.
An energetically most preferable state among those with solitons, polarons, 
bipolarons, and no localized excitations is not well understood~\cite{Bendikov,yamamoto2}. 
The stability of polarons against various perturbations such as an external electric field~\cite{Johansson1,Johansson2,Johansson3,Rakhmanova} 
and thermal noise~\cite{Ness,Liu2009} is still intensively studied even now, as is the dependence of
the polaron velocity on the applied field\cite{Horovitz,Johansson3,Xiaojing} and/or Coulomb interaction~\cite{Di}.

As of yet, only the destructive role of electric field and temperature on the stability of polarons 
has been studied. In this Rapid Communication, we report a {\it constructive role} of
current within the TLM model. Namely, we show that, at nonequilibrium steady states (NESS), current
induces polarons that are absent at equilibrium. 
It is a genuine nonequilibrium property, since there exists a lower threshold current below which the 
polaron cannot exist.
Although only the TLM model is discussed here, 
our analysis covers a wider class of systems.
This is because,
in the mean-field approximation, 
the TLM model is equivalent to 
the XXZ model, the extended Hubbard model
and the Gross-Neveu model in field theory~\cite{DHN,Brazavskii,CB1}.

The new polaron solutions are possible both for the spinful and spinless TLM models.
Thus, to emphasize the very role of the current, we discuss the {\it spinless} TLM model which is known to 
admit no polarons at equilibrium.
The self-consistent conditions out of equilibrium are derived based on the scattering-theoretical characterization of
NESS proposed by Ruelle\cite{Rue1,LecMath1880,JP1}. 

The Hamiltonian $H\equiv H_{S} + V+  H_B$ is composed of 
$H_S$ for the finite TLM chain, $H_B$ for the reservoirs, 
and $V$ for their interaction, which are given by
\begin{eqnarray}
\label{HamHS}
H_S &=&
\int_0^\ell dx
\Psi^\dag(x)
\left[-i \hbar v
\sigma_y\frac{\partial }{\partial x}+ 
\hat{\Delta} (x) \sigma_x
\right]
\Psi(x)
\nonumber \\
&&+
{1\over 2\pi \hbar v \lambda}\int_0^\ell dx 
\left[\hat{\Delta}(x)^2
+{1\over \omega_0^2}\hat{\Pi}(x)^2
\right]
\nonumber \\
V&=&\int d{\boldsymbol  k} \ 
\bigg\{ \hbar v_{\boldsymbol  k}
e^\dag(0) 
a_{{\boldsymbol  k}L} + \hbar
v_{\boldsymbol  k}
d^\dag(\ell) 
a_{{\boldsymbol  k}R} + (h.c.) \bigg\}
\nonumber \\
H_B &=& \integ{{\boldsymbol  k}}
\hbar( \omega_{{\boldsymbol  k}L} a_{{\boldsymbol  k}L}^{\dag}a_{{\boldsymbol  k}R} + 
\omega_{{\boldsymbol  k}R} a_{{\boldsymbol  k}R}^\dag a_{{\boldsymbol  k}R} ) \ ,
\end{eqnarray}
where $\Psi(x)=\left( d(x) , e(x)\right)^T$
is the two-component spinless fermionic field satisfying a boundary condition; $d(0)=0,\ e(\ell)=0$, $\hat{\Delta}(x)$
is the lattice distortion, $\hat{\Pi}(x)$ is the 
momentum conjugate to $\hat{\Delta}(x)$, $a_{\boldsymbol  k \nu}$ ($\nu=L,R$) 
are the annihilation 
operators for reservoir fermions with wave number ${\boldsymbol  k}$, 
$\hbar\omega_{{\boldsymbol  k}\nu}$ represents 
their energies measured from the zero-bias chemical potential at absolute zero 
temperature, $\sigma_x$ and $\sigma_y$ are the Pauli matrices, $\ell$ is the length of the system, 
$v$ is the Fermi velocity, $\lambda$ is the dimensionless coupling constant, 
and $\omega_0$ is the phonon frequency.
We assume that the coupling matrix elements $v_{\boldsymbol  k}$
as well as the density of states of the reservoirs are
energy independent\cite{footnote3}; thus,
the integral
$$
\frac{1}{i}\int d{\boldsymbol k} \frac{|v_{\boldsymbol k}|^2}
{\omega-\omega_{{{\boldsymbol k}}\nu}-i0}
\sim \pi\int d{\boldsymbol k} |v_{\boldsymbol k}|^2
\delta (\omega-\omega_{k \nu}),\ (\nu=L,R)
$$
becomes a positive constant $\Gamma$.

Next we describe
the mean-field approximation.
Since we are interested in NESS, the self-consistent condition is derived from the 
equation of motion for the lattice distortion,
\begin{eqnarray*}
\frac{\partial^2 \hat{\Delta}(x,t)}{\partial t^2}=
-\omega_0^2
\left\{\hat{\Delta}(x,t)+\pi \hbar v \lambda
\Psi^\dag(x,t) \sigma_x \Psi(x,t)
\right\} .
\end{eqnarray*}
Namely, the self-consistent equation is written
\begin{eqnarray}
\Delta(x)+\pi \hbar v \lambda
\langle\Psi^\dag(x,t) \sigma_x \Psi(x,t)\rangle^{MF}_\infty =0\ ,
\label{self}
\end{eqnarray}
where $\Delta(x)$ is the mean-field NESS average of $\hat{\Delta}(x)$, and
$\langle\cdots\rangle^{MF}_\infty$ represents the mean-field NESS average.
The mean-field NESS corresponds to the initial state where two reservoirs are 
in equilibrium with different chemical potentials, and 
it is characterized as a state 
satisfying Wick's theorem 
with respect to the {\it incoming} fields $\alpha_{{\boldsymbol  k}\nu}$ 
($\nu=L,R$) of the mean-field Hamiltonian,
and of having the two-point functions\cite{Buttiker,PTP}:
\begin{eqnarray*}
\langle \alpha_{{\boldsymbol  k}\nu}^\dag \alpha_{{\boldsymbol  k}'\nu}
\rangle_\infty&=& f_\nu(\hbar\omega_{k \nu}) \delta({\boldsymbol  k}-{\boldsymbol  k}')\ ,\ 
\ (\nu=L,R)
\end{eqnarray*}
where $\alpha_{{\boldsymbol  k}\nu}$ corresponds to the unperturbed field 
$a_{{\boldsymbol k}\nu}$,
$f_\nu(x)\equiv 1/(\exp\{(x-\mu_\nu)/T\}+1)$
is the Fermi distribution function with temperature $T$ and, 
chemical potential $\mu_L=-eV/2$ and $\mu_R=eV/2$ 
(the Boltzmann constant is set to be unity).

\begin{figure}
 \rotatebox[origin=c]{-90}{\includegraphics[width=30mm]{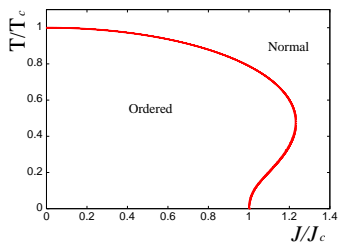}}
\vskip -4mm
\caption{\label{Phase diagram} Phase diagram on the $J$-$T$ plane.}
\end{figure}

\begin{figure}[htbp]
 \begin{minipage}{0.49\hsize}
 \begin{center}
 \rotatebox[origin=c]{-90}{\includegraphics[width=30mm]{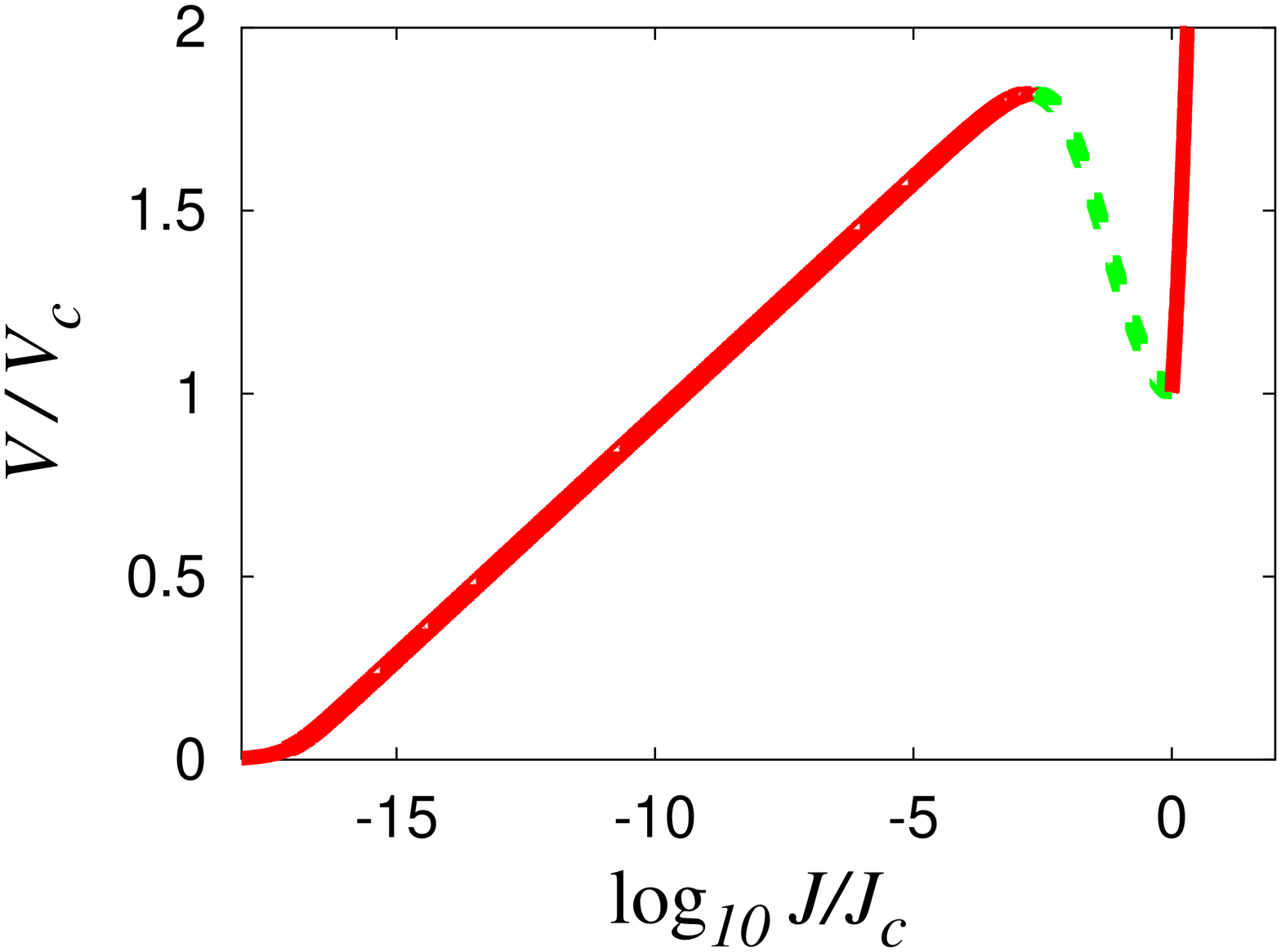}}
 \end{center}
 \end{minipage}
  \begin{minipage}{0.49\hsize}
  \begin{center}
 \rotatebox[origin=c]{-90}{\includegraphics[width=30mm]{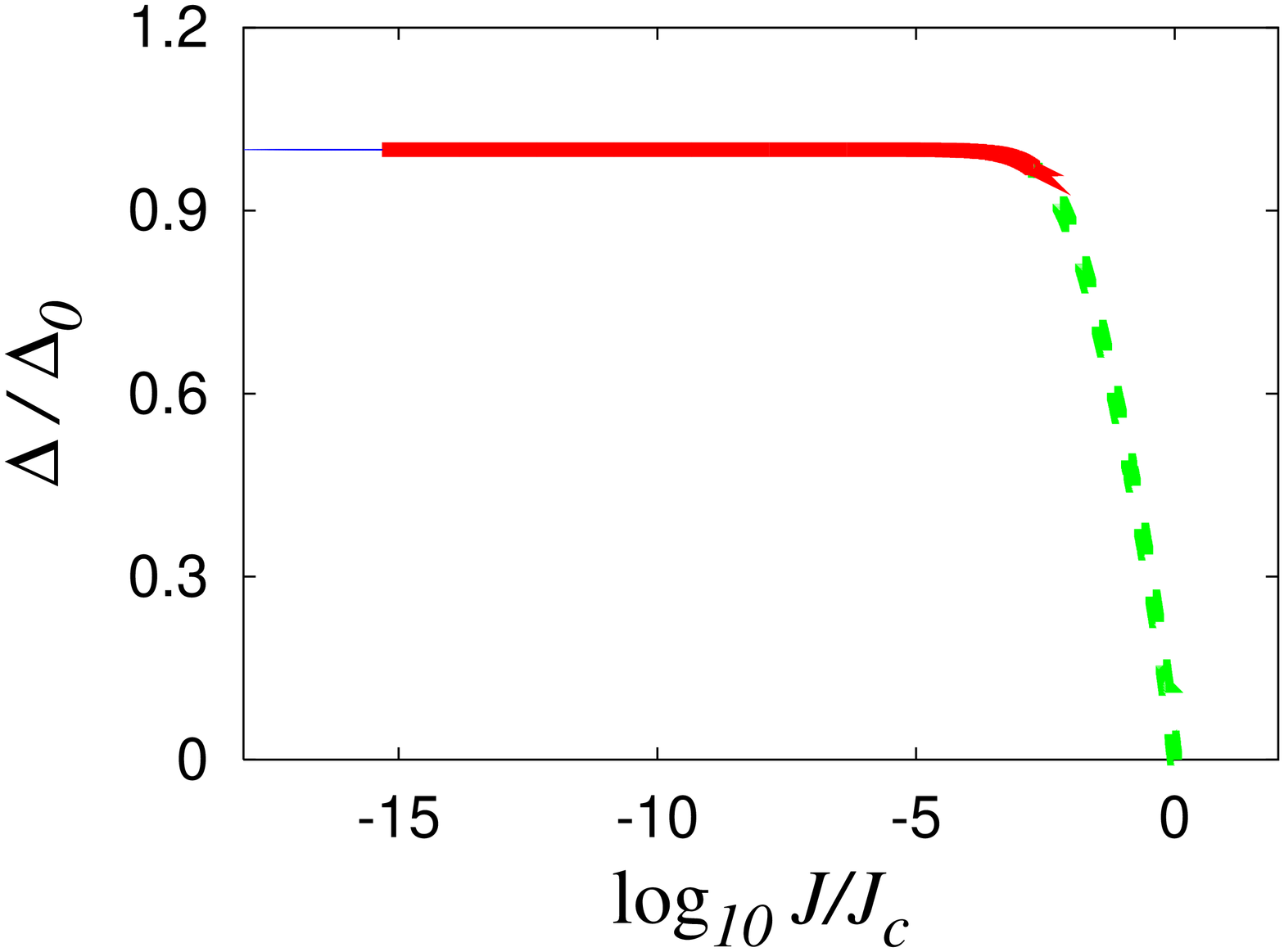}}
  \end{center}
 \end{minipage}
\vskip -4mm
  \caption{
The left figure shows current-voltage characteristics at $T=0.05T_c$. 
The right figure shows the current dependence of $|\Delta_0|$ at $T=0.05T_c$.
In these figures, the solid line is stable, and the dashed line is stable only at a constant current.
In the right figure, only the bold solid line admits polarons.
}  \label{uniform}
\end{figure}

At first, we briefly review the previous results on
the uniformly dimerized case\cite{PTP}, in which the average lattice distortion is constant: $\Delta(x)=\Delta_0$, 
the fermionic spectrum has a gap $2|\Delta_0|$, and $\Delta_0$ obeys
the gap equation
\begin{eqnarray}
\int_{\Delta_0/\hbar}^{\omega_c}d\omega
\sum_{\nu=L,R} 
\frac{f_\nu(-\hbar\omega)-f_\nu(\hbar\omega)}
{\sqrt{(\hbar\omega)^2-\Delta_0^2}}
=\frac{2}{\hbar\lambda}\ ,
\label{self2}
\end{eqnarray}
where $\omega_c$ is the energy cut-off. 
This reduces to a well-known 
expression at equilibrium in the absence of a bias voltage.
This equation is valid when the chain length $\ell$ is sufficiently long.
The average lattice distortion $\Delta_0$ is found to be a multi-valued function of the bias voltage when $T<T^*\sim 0.5571\times T_{c}$. But,
in terms of the current, which is given by
\begin{eqnarray}
J
=\frac{G_0}{e}
\int_{|\Delta_0|<|\epsilon|<\hbar\omega_c} d\epsilon 
\frac{\sqrt{\epsilon^2- \Delta_0^2}}
{|\epsilon|}
\left[f_R(\epsilon)-f_L(\epsilon)\right]
\ ,
\label{current}
\end{eqnarray}
it is a single-valued function at every temperature.
In the above, $G_0={e^2v\Gamma}/\{\pi\hbar(v^2+\Gamma^2)\}$ is 
the conductance in the normal phase.
Thus, the temperature and the current are chosen as control
parameters.
The phase diagram on the $J$-$T$ plane and the current dependence of the
average lattice distortion are shown, respectively, in Fig.~\ref{Phase diagram}
and Fig.~\ref{uniform} (left), for $\lambda^{-1}=2.4$. In these figures, the average lattice distortion,
the temperature, and the current are scaled, 
respectively, by the zero-bias lattice distortion $\Delta_{c}\equiv\hbar\omega_c/\cosh\lambda^{-1}$
at $T=0$, the zero-bias critical temperature 
$T_{c}\equiv 2\hbar\omega_c \exp(\gamma-\lambda^{-1})/\pi$ ($\gamma$: Euler constant), and the 
critical current $J_{c}\equiv G_0 V_c$ at $T=0$, where 
$V_{c}\equiv 2\hbar\omega_c\exp(-\lambda^{-1})/e$ is the critical bias voltage at $T=0$.
The multi-valued property of the average lattice distortion with respect to the voltage results 
in NDC for $T<T^*$.
We note that NDC was reported in the field-driven SSH model\cite{Wei}, as well as in the stochastically-driven XXZ and extended Hubbard models\cite{GiulianoBenenti2009,GiulianoBenenti2009a}.
However, we also note that the NDC reported in these models 
is different from the one discussed here.
Since, as detailed above,
the current is a multi-valued function of the chemical potential difference. 
This point merits further investigation.

Next we investigate the solitons and polarons. 
Observing that the only difference between the NESS and equilibrium cases is 
that the Fermi distribution is replaced with the averaged distribution $\{f_L(\epsilon)+f_R(\epsilon)\}/2$,
the self-consistent Eq.~(\ref{self}) is expected to have similar solutions 
to the case at equilibrium. 
It is easy to verify that Eq.(\ref{self}) admits a soliton solution\cite{preprint} similar to
that of the equilibrium case\cite{TLM,Brazavskii}
\begin{eqnarray*}
\Delta(x)=\Delta_0 \tanh \kappa_s (x-a),\ \  \kappa_s=\Delta_0/(\hbar v)\ ,
\end{eqnarray*}
where the amplitude $\Delta_0$ is the solution of the 
gap equation (\ref{self2})
for the uniformly dimerized phase, and $a={\rm O}(\ell)$ represents the center of the soliton.
At the same time, a midgap state appears with energy $\hbar\omega =0$ in the fermionic spectrum.
Note that, even when solitons exist, the current is still 
given by (\ref{current}). 
Then, following Brazovskii-Kirova\cite{BzCB1} and Campbell-Bishop \cite{BzCB2}, we look for a static polaron solution of the following form
\begin{eqnarray*}
\Delta(x)&=&
\Delta_0-\hbar v\kappa_0 (t_+ -t_-)
\\
t_\pm &\equiv& \tanh \kappa_0(x-a\pm x_0)\ ,\ \ 
\tanh 2\kappa_0 x_0=\frac{\hbar v\kappa_0}{\Delta_0}\ ,
\nonumber
\end{eqnarray*}
where $\Delta_0$ and $x_0$ are parameters that are determined self-consistently, and
$a$ is the position of the polaron center on the order of $\ell$.
As in the equilibrium case, the corresponding fermionic spectrum consists of continuum states 
with energy $\hbar\omega=\pm\sqrt{(\hbar vk)^2+\Delta_0^2}$ ($|k|<\omega_c/v$), and 
midgap states with
energies $\hbar\omega=\pm\sqrt{\Delta_0^2-(\hbar v\kappa_0)^2}\equiv\pm\hbar\omega_B$.
Even though the coupling between the midgap states and the reservoirs is exponentially small for long chain length $\ell$, it still controls the occupation of
the midgap states at NESS\cite{footnote4}.
Therefore, one should carefully take a long chain limit, resulting in a self-consistent equation~(\ref{self})
\begin{eqnarray*}
&&\mskip -15 mu I_{B}+I_{S}=
-\frac{\Delta(x)}{\hbar v\lambda}
\\
&&\mskip -15 mu I_{B} \equiv -
\frac{\pi\omega_B}{4 v}(t_+ - t_-)
\frac{\sinh\hbar\beta\omega_B}{\cosh\hbar\beta\omega_B +\cosh\frac{\beta eV}{2}}
\\
&&\mskip -15 mu
I_{S} \equiv 
-\int_{|\Delta_0|/\hbar}^{\omega_c}\mskip -5 mu d\omega
\frac{\omega^2\Delta(x)-\omega_B^2\Delta_0}
{2 \hbar v^2 \kappa \left(\omega^2 -\omega_B^2 \right)}
\frac{\sinh\hbar\beta\omega}{\cosh\hbar\beta\omega+\cosh\frac{\beta eV}{2}},
\end{eqnarray*}
where $\beta=1/T$, 
$I_S$ is a contribution from the continuum states 
, and $I_B$ is a contribution from the midgap states with energy 
$|\hbar\omega|<|\Delta_0|$ (see \cite{preprint}).
Comparing term by term, the gap equation~(\ref{self2}) is obtained, and the equation for energies $\pm\hbar\omega_B$ of the midgap states
\begin{eqnarray}
&&\int_{|\Delta_0|/\hbar}^{\omega_c}
\frac{\omega_B \ d\omega}{\sqrt{\omega^2-\Delta_0^2/\hbar^2}\left(\omega^2 -\omega_B^2 \right)}
\frac{\sinh\hbar\beta\omega}{\cosh\hbar\beta\omega+\cosh\frac{\beta eV}{2}}
\nonumber
\\
&&\mskip 100 mu =\frac{\pi}{2v\kappa_0}
\frac{\sinh\hbar\beta\omega_B}{\cosh\hbar\beta\omega_B+\cosh\frac{\beta eV}{2}}
\label{self1}
\end{eqnarray}

\begin{figure}[htbp]
 \begin{minipage}{0.49\hsize}
 \begin{center}
 \rotatebox[origin=c]{-90}{\includegraphics[width=30mm]{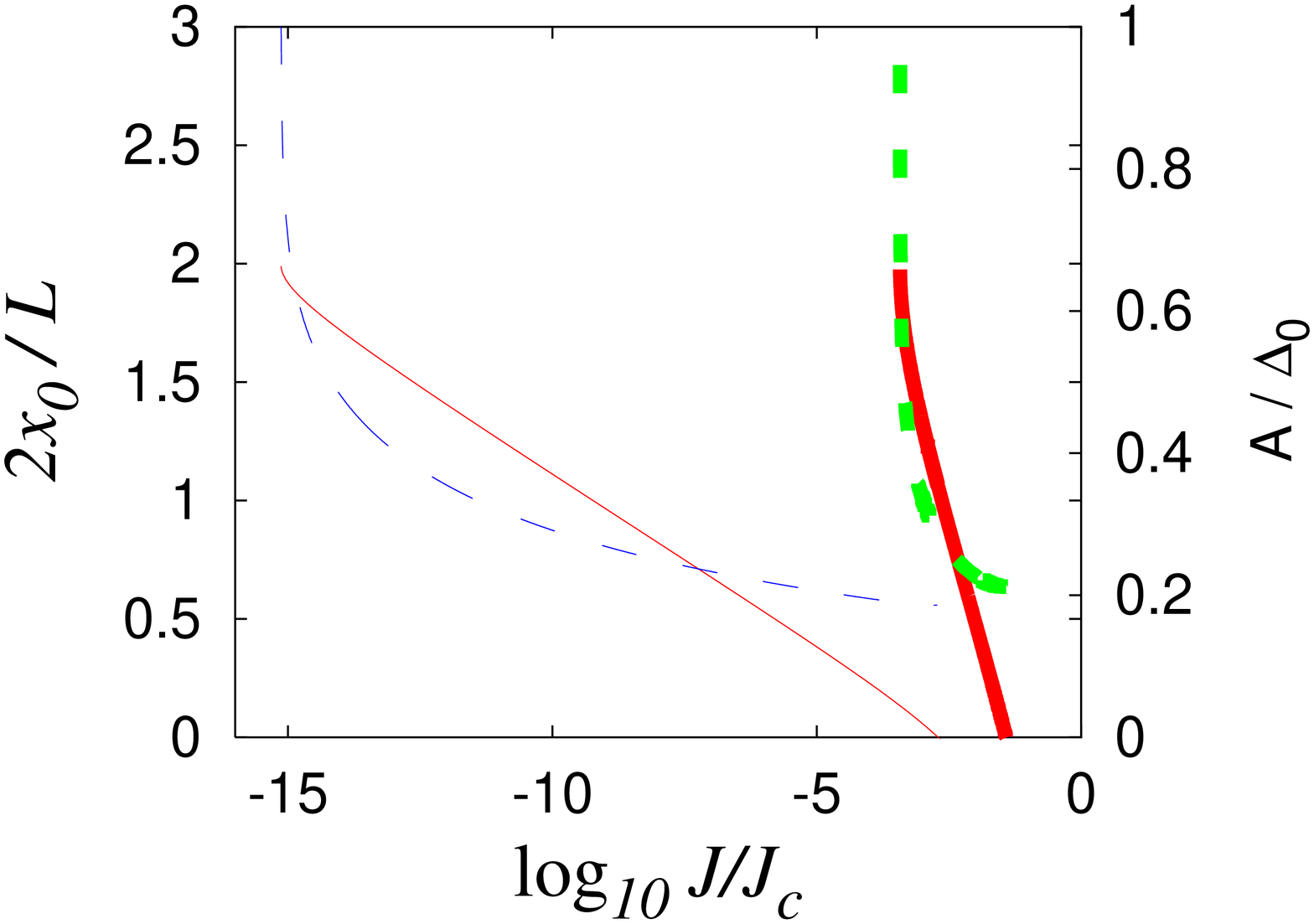}}
 \end{center}
 \end{minipage}
  \begin{minipage}{0.49\hsize}
  \begin{center}
 \rotatebox[origin=c]{-90}{\includegraphics[width=30mm]{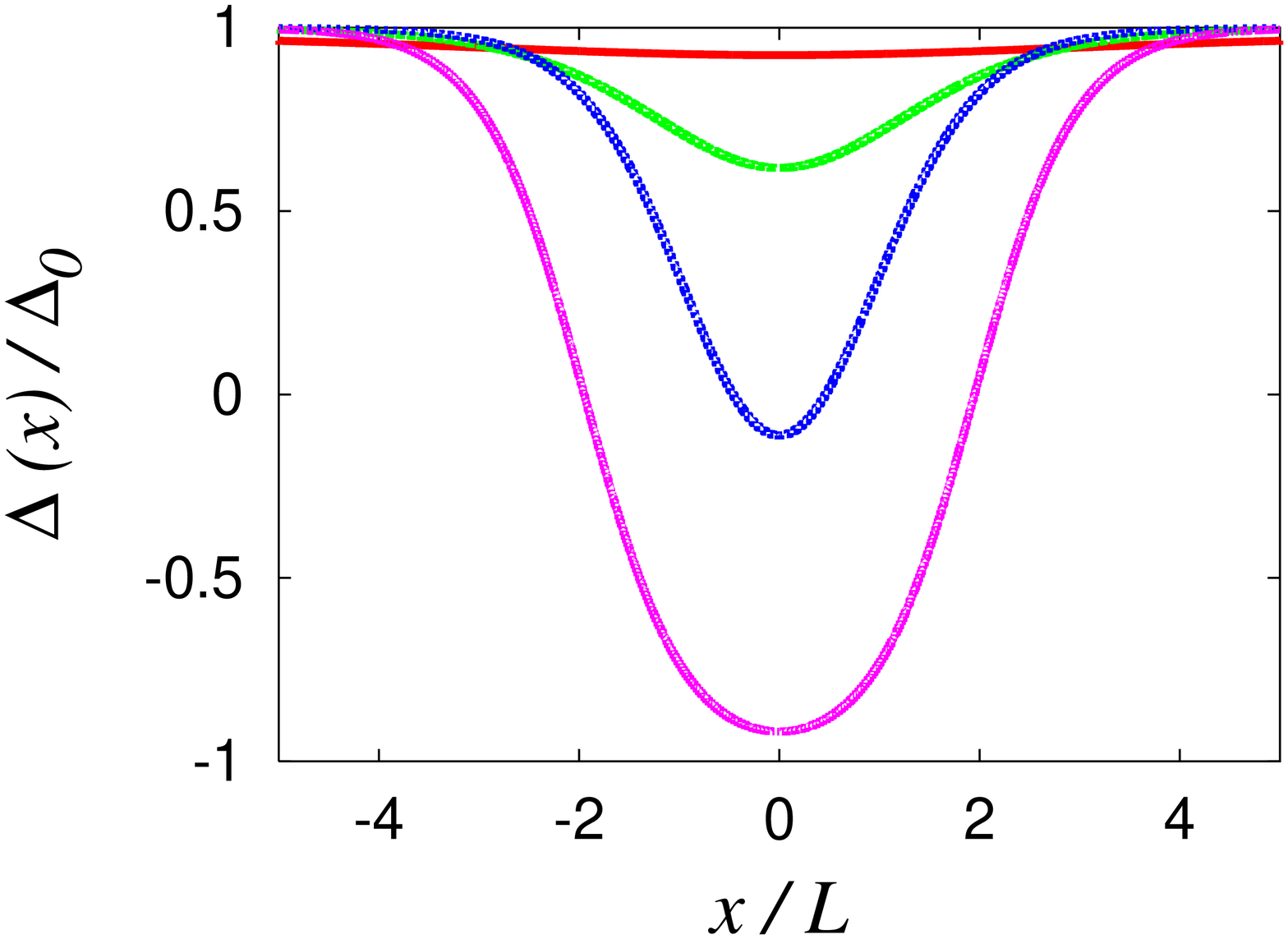}}
  \end{center}
 \end{minipage}
\vskip -4mm
 \caption{The left figure shows the current dependence of the amplitude $A$ (the solid lines) and the soliton size $2x_0$ (the dashed lines) at $T=0.05\times T_{c}$ (the thin lines) 
and $T=0.2\times T_{c}$ (the bold lines).
$2x_0$ is scaled by $L=\hbar v/\Delta_c$.
The right figure shows a typical lattice profile at $T=0.05\times T_{c}$. 
From top to bottom, $J=10^{-3}J_c,\ 10^{-5}J_c,\ 10^{-10}J_c$ and $10^{-15}J_c$}\label{profile}
\end{figure}

\begin{figure}
 \rotatebox[origin=c]{-90}{\includegraphics[width=30mm]{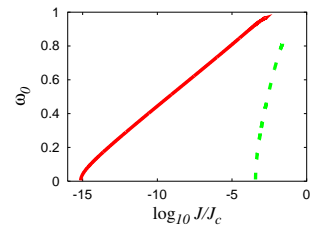}}
\vskip -4mm
\caption{\label{bound}Current dependence of the positive-bound-state energy $\hbar\omega_B$ at $T=0.05\times T_{c}$ (the solid line) and $T=0.2\times T_{c}$ (the dashed line).}
\end{figure}
Note that the current is still given by (\ref{current}) for the polaron solutions.
Eqs.(\ref{self2}) and (\ref{self1}) have a nontrivial solution
{\it only when} the current (equivalently, the bias voltage) lies between
the lower and upper threshold values $J_1(T)<J<J_2(T)$ ($V_1(T)<V<V_2(T)$), and
the temperature is lower than $T^*$, under which the system shows NDC.
As seen in the left figure of Fig.~\ref{profile}, the polaron width $2x_0$ and amplitude 
$A\equiv 2(\hbar v\kappa_0)^2/(|\Delta_0|+\hbar\omega_B)$ are decreasing functions of the current.
When the current (equivalently, the bias voltage) 
approaches the lower threshold 
$J_1(T)$ ($V_1(T)$), the polaron width diverges and the polaron amplitude 
approaches the soliton amplitude $2|\Delta_0|$. This indicates that 
the polaron splits into a soliton-antisoliton pair. On the other hand, when the current 
(the bias voltage) approaches the upper threshold $J_2(T)$ ($V_2(T)$), both the width and
amplitude of the polaron vanish, and the polaron solution reduces to the uniform solution. 
Typical profiles of the polaron solution are shown in the right figure of Fig.~\ref{profile}.

As mentioned above, $|\Delta_0|$ is a multi-valued function of the bias voltage
and, for a given voltage, several uniform phases are possible. Although this suggests
the possibility that collective local excitations can separate uniform domains with different values of
$|\Delta_0|$, there exist only those interpolating uniform phases with the {\it same} $|\Delta_0|$, such 
as the solitons and polarons just discussed. This is because charge conservation implies that the current $J$ remains constant over the chain, and $\Delta_0$ is a single-valued function of $J$.
Also, it is interesting to note that the existence of the polaron solution is 
related to the linear stability studied previously\cite{PTP}. 
Indeed, the polaron solution exists when the uniform phase with $\Delta_0$ is stable {\it both} 
at constant current and constant bias voltage (the solid curves in the right figure of Fig~\ref{uniform}), but it does not exist if the uniform phase is unstable at constant voltage 
(the dashed curve in the right figure of Fig~\ref{uniform} ).
Because of this property, 
there is one-to-one correspondence between the current and bias voltage intervals where the polaron solution is possible, $J_1(T)<J<J_2(T)$ and $V_1(T)<V<V_2(T)$, respectively. 
This aspect and the non-existence of the polaron solution for $T>T^*$
deserve further investigation.
Note that the states on the thin solid curve do not admit polaron solutions.

The possibility of the polaron solution at NESS can be qualitatively
understood as follows. Recall that the polaron at equilibrium is possible only in the spinful case. 
With the corresponding fermionic state, the lower midgap state is occupied 
by two fermions with opposite spins, and the upper midgap state is occupied by an unpaired fermion.
In the half-filled spinless case at equilibrium, such an asymmetric occupation is not possible.
This seems to suggest the necessity of the particle-hole symmetry breaking for the polaron formation.
This seems to suggest that it is necessary for the particle-hole symmetry to break for polaron formation.
In contrast, at NESS, the particle-hole symmetry is broken by the bias voltage even for the half-filled spinless case.
This is because the fermionic occupation is controlled by 
$(f_L(\epsilon)+f_R(\epsilon))/2$, which is not symmetric under the exchange of particles and holes.

It is interesting to note that, at low temperatures, the width $J_2(T)-J_1(T)$ of the current interval that admits polaron solutions
increases with an increase in temperature, while the width $V_2(T)-V_1(T)$ of the voltage interval decreases with temperature.
These behaviors of the current and voltage are consistent, because
the phases admitting polaron solutions tend to become insulating phases as $T\to 0$, which implies $\lim_{T\to 0}(J_2(T)-J_1(T))/(V_2(T)-V_1(T))=0$; thus, the decrease of $V_2(T)-V_1(T)$ with an increase of $T$ does not contradict the increase of $J_2(T)-J_1(T)$.
Because of the discontinuity at $T=0$ of the R.H.S. of Eq.~(\ref{self1}), which behaves like the Fermi distribution function, absolute zero temperature is a singular point. Indeed, 
at $T=0$, Eq.~(\ref{self2}) and Eq.~(\ref{self1}) admit a polaron solution with $\hbar\omega_B=(\pi^2/16+1)^{-1/2}\Delta_c$
only when\\
$V=\{(\pi^2/16+1)\cosh^2 \lambda^{-1}\}^{-1/2} \exp(\lambda^{-1}) V_c$ and
$J=0$.

The existence of solitons and polarons has been verified by spectroscopic experiments, where the energies of the associated midgap states are observed\cite{SSHrev,OH2,Polaronex}.
Fig.~\ref{bound} shows the current dependence of the energy $\hbar\omega_B$ for the midgap state at $T=0.05\times T_{c}, 0.2\times T_{c} \ (<T^*)$.
As shown in the figure, $\hbar\omega_B$ is a monotonically increasing function of the current, and it approaches 0 for $J\to J_1(T)$; and $\Delta_0$ for $J\to J_2(T)$; this reflects the change of the polaron profile.
Polarons in a {\it spinful system} possess this same feature, since the corresponding self-consistent equation is obtained simply by replacing $\lambda$ in (\ref{self2}) with $2\lambda$.
Namely, the energies $\pm\hbar\omega_B$ of the midgap states associated with NESS polarons change
from 0 to $\pm|\Delta_0|$ as the current increases, while those with equilibrium polarons in a {\it spinful} system are fixed at $\pm\hbar\omega_B=\pm |\Delta_0|/\sqrt{2}$. 
Such a current-induced shift of energy spectra might be observed by spectroscopic experiments.

In summary, we have studied solitons and polarons in the open spinless TLM model, and in particular, we have shown that polarons are possible {\it only out of equilibrium}. The polaron formation is a genuine nonequilibrium phenomenon, as there exists a lower critical current $J_1(T)$ (equivalently, a lower critical bias voltage $V_1(T)$), below which polarons are not possible. 
This observation suggests that the new polaron is an example of microscopic dissipative structure. 
Also, we have shown that the critical temperatures for polaron formation and the appearance of the negative differential conductivity (NDC) agree, although polarons are not 
allowed at the current found in the NDC regime.
The energies of the midgap states associated with polarons are shown to crucially depend on the current, which might be observed by spectroscopic experiments.


The authors thank
T. Prosen, T. Monnai, N. Weissburg, and T. S. Evans for fruitful discussions.
This work is partially supported by Grants-in-Aid for Scientific
Research (Nos. 17340114, 17540365 and 21540398), and by the ``Academic Frontier'' Project from MEXT.

\newpage 
\bibliography{aps11d}

\begin{thebibliography}{37}
\bibitem[{foo({\natexlab{a}})}]{footnote3}
\bibinfo{note}{To derive the gap equation, this assumption can be replaced by
  weaker one we used in \cite{PTP}.}
\bibitem[{foo({\natexlab{c}})}]{footnote4}
\bibinfo{note}{Due to the exponentially small coupling between the midgap
  states and the reservoirs, $I_B$ depends on the position of polarons' center
  $a$ if one consider NESS driven by reservoirs with different temperatures.}
\end{thebibliography}

\end{document}